\documentclass{mn2e}
\usepackage{graphicx}

\def\laeq{\stackrel{<}{\scriptstyle \sim}}
\title [An Asymmetric Dust Shell] {The Case for Asymmetric Dust
Around a C-Rich AGB Star} 
\author [M.W.Feast, P.A.Whitelock \& F.Marang] 
{Michael W. Feast$^{1}\thanks{e-mail: 
mwf@artemisia.ast.uct.ac.za}$, Patricia A. Whitelock$^{2}$, and Freddy
Marang$^{2}$\\
$^{1}$ Astronomy Department, University of Cape Town, 7701, Rondebosch, South
Africa.\\
$^{2}$ South African Astronomical Observatory, P.O.Box 9, 7935, Observatory,
South Africa.}
\begin{document}
\maketitle
\begin{abstract}
   $JHKL$ observations of the mass-losing carbon Mira variable IRAS\,15194--5115
(II\,Lup) extending over about 18 years are presented and discussed. 
The pulsation period is 575 days and has remained essentially constant over
this time span. The star has undergone an extensive obscuration minimum
during this time. This is complex and, like 
such minima in similar objects, e.g. R\,For,
does not fit the model predictions of a simple long term periodicity. Together
with the high resolution observations of Lopez et al. the results
suggest that the obscuration changes are due to the formation of dust clouds
of limited extent in the line-of-sight. This is an RCB-type model. 
The effective reddening law at $J$ and $H$ is similar to that found
for R\,For. 
\end{abstract}
\section{Introduction}
  IRAS\,15194--5115 (II\,Lup) was classified as a carbon star by Epchtein et al.
(1987) and Meadows et al. (1987). 
At $12\mu\rm m$ it is the third brightest carbon star in the sky.
It is a Mira variable for which
Le Bertre (1992) gave an approximate period of 580 days. The circumstellar
environment of the star has been studied by a number of workers. For 
instance Ryde et al. (1999) modelled the radio and far infrared rotational
lines of CO, whilst Groenewegen et al. (1998) modelled the infrared
spectral energy distribution. The star has similarities to the proto-type
mass-losing carbon Mira IRC+10\,216 (CW\,Leo) which has a pulsation period
of 630 days. Ryde et al. derive a mass-loss rate of 
$1 \times 10^{-5}\rm M_{\odot} yr^{-1}$ for II\,Lup and suggest that the wind 
characteristics of the object have not changed over the past few thousand
years. However, they note that their data do not preclude variations in the
mass-loss rate in the inner parts of the circumstellar material by a
factor of three. Both Ryde et al. and Groenewegen et al. find that the
dust-to-gas ratio in II\,Lup is about twice that in the envelope of CW\,Leo.
However, the main difference between II\,Lup and CW\,Leo is in the
$\rm ^{12} C / ^{13} C$ ratio. Ryde et al. estimate this as 5.5 for II\,Lup.
Whereas Kahane et al. (1992) found $44 \pm 3$ for CW\,Leo. One
possibility, suggested by Ryde et al. for the relatively low ratio in
II\,Lup, is that the star has recently gone through a period of hot-bottom
burning.

Drops in the near infrared brightness of a carbon Mira on a time-scale long 
compared
with the pulsation period, were first seen in 
the Galactic Mira R\,For (Feast et al. 1984). In that case a decline at
visual wavelengths was also reported. These events were attributed to
increased absorption in the line-of-sight.
Further observations of R\,For were published by
Le Bertre (1992). The most extensive investigation of this phenomenon
is that of Whitelock et al. (1997) who discuss the near-infrared
light curves of 11 large amplitude carbon variables from data covering
time periods of from 9 to 22 years together with visual observations
for some of them. 

Here we present and discuss $JHKL$ (1.2, 1.6, 2.2 and 3.5
$\mu$m) observations  of II\,Lup over a time period of about 18 years.
The results are of particular interest for at least three reasons. (1) 
During this time period the object underwent an extended obscuration
phase. (2) The pulsation variations are well covered, early and late in the
time period involved. (3) At one point during the obscuration phase,
infrared speckle observations (Lopez et al. 1993) provide estimates of
the relative contributions of the starlight and the dust-shell emission
in the infrared.  These estimates are helpful in
interpreting the data presented here.

\section{Observations}
 The SAAO near infrared photometry of II\,Lup is listed in Table 1.
The times of the observations are given as Julian Date (JD) from which
2440000 has been subtracted.
The $JHKL$ measurements were all made with the
MKII photometer on the 0.75m telescope at SAAO, Sutherland. They are on
the current SAAO system as defined by the standards of Carter (1990).
The photometry is accurate to better than $\pm 0.03$mag at $JHK$ and
to better than $\pm 0.05$mag at $L$. Three of the early measurements were
published by Lloyd Evans \& Catchpole (1989). This is the star called 
WO48 in that paper, but note that two of the JDs were inadvertently
transposed there.

\begin{table}
\begin{center}
\caption[]{\label{table.JHKL} SAAO Near-infrared photometry for II~Lup}
\begin{tabular}{rcccr}
\hline
\multicolumn{1}{c}{(JD)}& $J$ & $H$ & $K$ & \multicolumn{1}{c}{$L$} \\
\multicolumn{1}{c}{--2440000}& \multicolumn{4}{c}{(mag)}\\
\hline
 5938.26&   5.95&   3.40&   1.46&  --0.80\\
 6309.28&   7.90&   4.97&   2.66&   0.04\\
 7214.61&   7.32&   4.34&   2.01&  --0.69\\
 7240.54&   7.43&   4.42&   2.09&  --0.60\\
 7263.51&   7.58&   4.56&   2.21&  --0.52\\
 7280.46&   7.72&   4.70&   2.31&  --0.38\\
 7329.37&   8.16&   5.11&   2.65&  --0.10\\
 7356.32&   8.43&   5.34&   2.87&   0.05\\
 7392.25&   8.60&   5.56&   3.07&   0.18\\
 7608.43&   8.22&   5.24&   2.74&  --0.15\\
 7686.40&   7.76&   4.79&   2.34&  --0.44\\
 7720.27&   7.66&   4.67&   2.24&  --0.54\\
 7743.25&   7.60&   4.64&   2.24&  --0.52\\
 7989.57&   8.72&   5.78&   3.27&   0.28\\
 8093.28&   8.64&   5.76&   3.28&   0.38\\
 8380.50&   7.47&   4.60&   2.21&  --0.53\\
 8434.33&   7.72&   4.84&   2.44&  --0.39\\
 8458.29&   7.85&   4.99&   2.59&  --0.27\\
 8706.54&   8.08&   5.42&   3.07&   0.18\\
 8748.53&   7.71&   5.02&   2.69&  --0.08\\
 8793.32&   7.27&   4.58&   2.30&  --0.44\\
 9003.55&   7.24&   4.48&   2.22&  --0.48\\
 9168.32&   8.23&   5.41&   3.11&   0.32\\
 9236.26&   8.26&   5.40&   3.07&   0.28\\
 9502.44&   7.51&   4.52&   2.16&  --0.61\\
 9589.25&   8.26&   5.15&   2.67&  --0.24\\
 9831.59&   9.21&   6.09&   3.46&   0.39\\
10234.38&   8.97&   5.71&   3.12&   0.12\\
10256.37&   9.09&   5.82&   3.22&   0.27\\
10478.60&   8.35&   5.17&   2.63&  --0.23\\
10503.58&   8.18&   5.01&   2.49&  --0.37\\
10589.48&   7.78&   4.61&   2.14&  --0.60\\
10618.35&   7.75&   4.58&   2.11&  --0.62\\
10802.59&   8.24&   5.21&   2.72&  --0.05\\
10995.51&   7.92&   5.07&   2.71&   0.07\\
11053.51&   7.43&   4.67&   2.38&  --0.25\\
11235.48&   6.71&   4.04&   1.84&  --0.71\\
11300.46&   6.71&   4.11&   1.98&  --0.53\\
11352.43&   6.85&   4.26&   2.17&  --0.23\\
11417.23&   6.95&   4.39&   2.35&   0.03\\
11613.54&   6.39&   4.00&   2.04&  --0.22\\
11678.41&   5.88&   3.56&   1.68&  --0.59\\
11712.35&   5.56&   3.28&   1.49&  --0.71\\
11747.26&   5.42&   3.13&   1.39&  --0.79\\
11782.26&   5.48&   3.17&   1.41&  --0.75\\
11963.57&   6.18&   3.81&   1.97&  --0.15\\
11979.57&   6.22&   3.85&   2.02&  --0.10\\
12033.47&   6.32&   3.95&   2.13&   0.01\\
12068.40&   6.35&   4.00&   2.18&   0.12\\
12116.33&   6.34&   3.98&   2.18&   0.09\\
12128.29&   6.32&   3.97&   2.14&   0.07\\
12159.24&   6.09&   3.79&   1.99&  --0.06\\
12178.22&   5.80&   3.54&   1.78&  --0.26\\
12322.59&   5.29&   3.04&   1.35&  --0.76\\
12324.62&   5.30&   3.05&   1.34&  --0.79\\
12347.60&   5.38&   3.15&   1.46&  --0.74\\
12382.50&   5.40&   3.11&   1.39&  --0.69\\
12426.33&   5.56&   3.24&   1.48&  --0.61\\
12457.31&   5.69&   3.34&   1.59&  --0.49\\
12524.23&   6.24&   3.79&   1.94&  --0.14\\
12690.64&   6.52&   4.00&   2.14&   0.04\\
\hline
\end{tabular}
\end{center}
\end{table}

\section{Discussion}
\subsection{The Light Curves}
\begin{figure*}
\includegraphics[height=20cm]{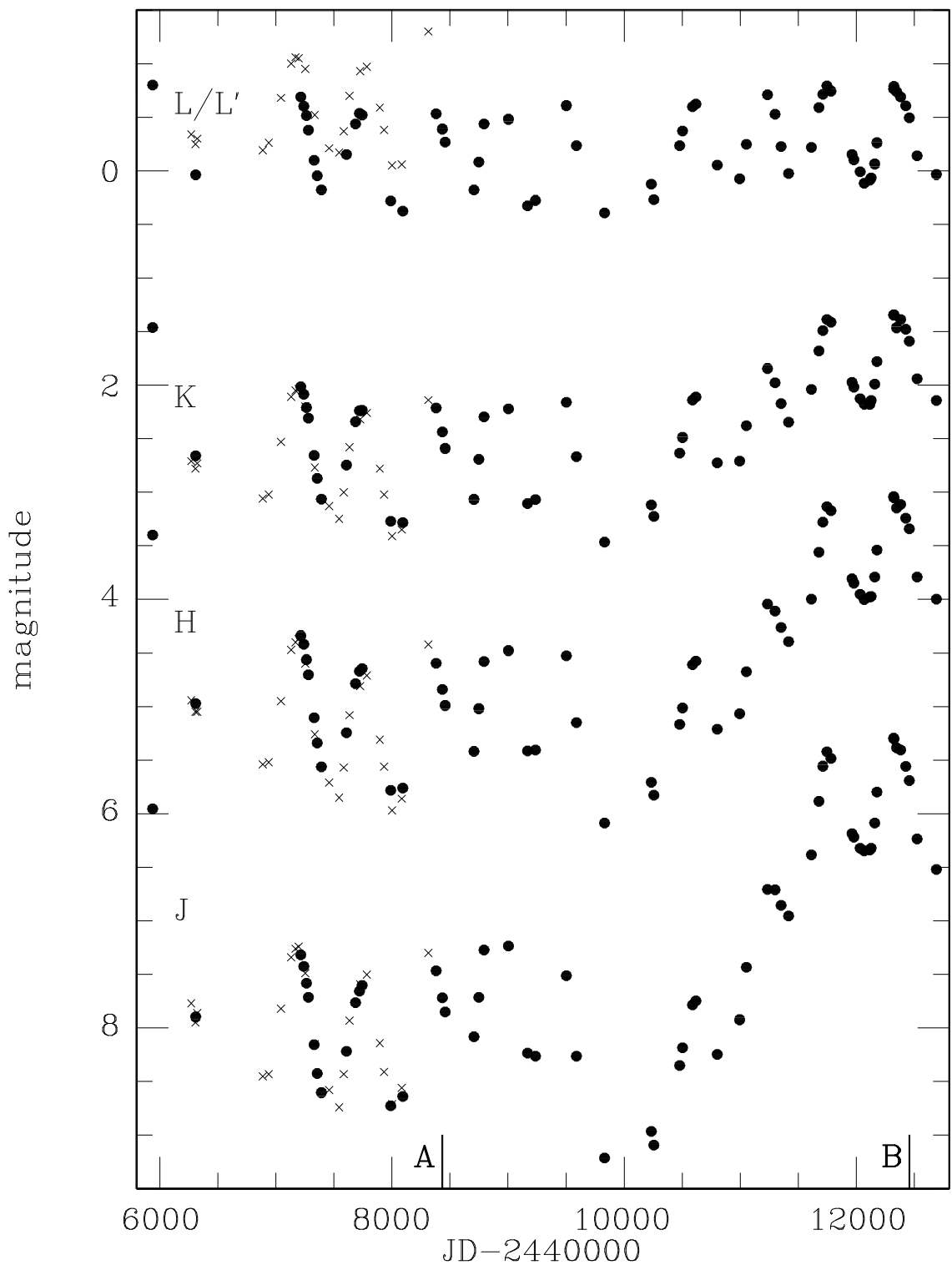}
\caption{A plot of the $JHKL$ magnitudes of II\,Lup against JD. Non-SAAO
observations are shown as crosses; see text for details}
\end{figure*}

Figure 1 shows the $JHKL$ light curves  The figure
also includes as crosses the measurements from Le Bertre (1992) as well as
the observations from Epchtein et al. (1987), Meadows et al. (1987)
and Groenewegen et al. (1993). There are likely to be differences between
the various photometric systems.
This is clearly seen  at $L$ where an $L^{\prime}$ filter was used by the ESO
observers.
There is no clear evidence
of a significant difference between the SAAO $J$ data and that of the other
observers. In any case the non-SAAO data are not used except to 
define the general shape of the light variations. The very bright $L^{\prime}$ 
magnitude at JD\,2448314 is from Groenewegen et al. (1993). 
Their $JHK$ measures do not appear unusual. They only report this single
observation and 
we have not considered their
$L^{\prime}$ measurement in the later discussion.
\begin{figure*}
\includegraphics[width=17.5cm]{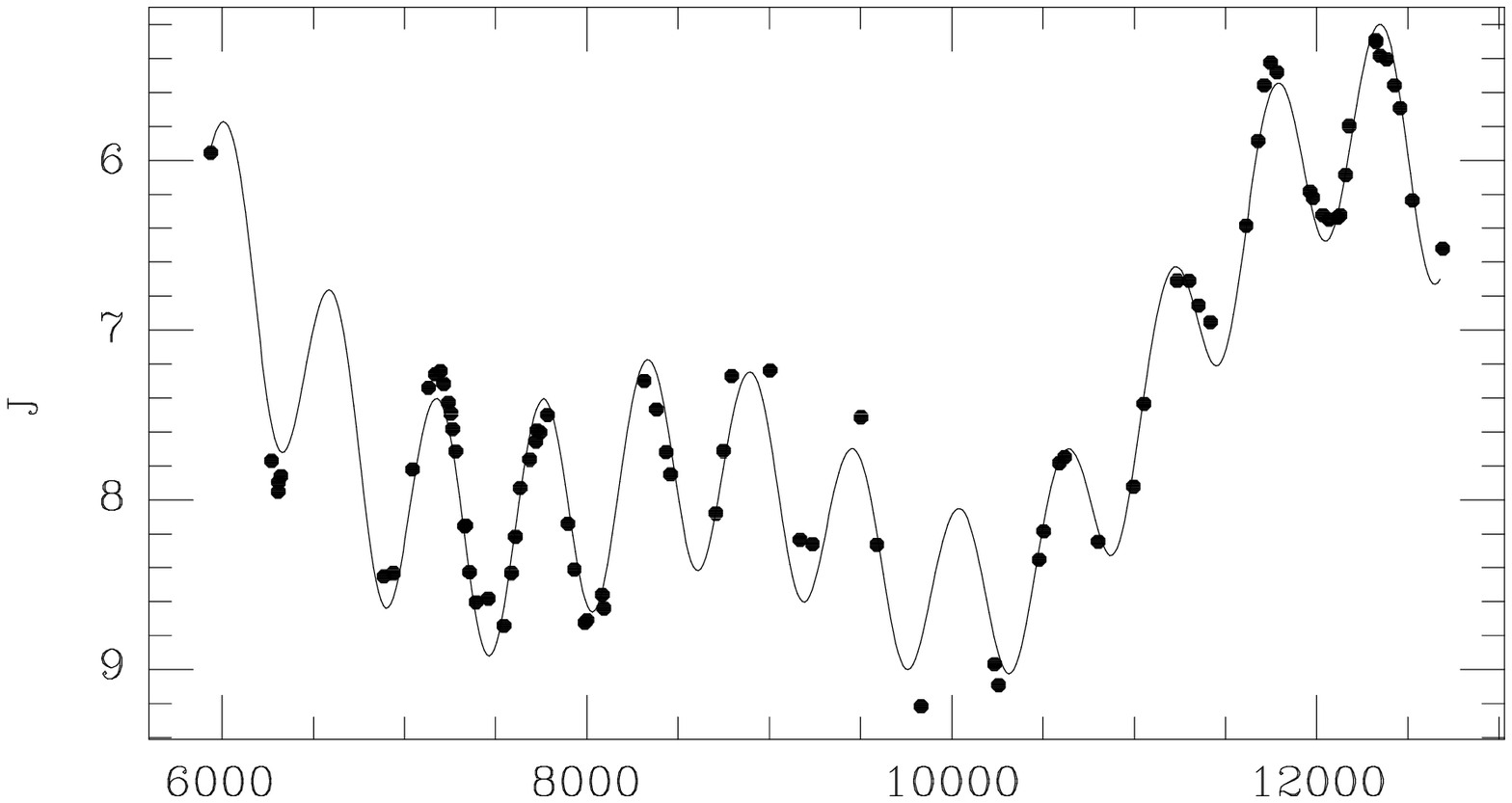}
\caption{The $J$ magnitudes of II\,Lup plotted against JD--2440000 with a 
curve showing the combination of periods of 575 and 6900 days and their first 
harmonics.}
\end{figure*} 

In the following we concentrate on discussing the $J$ and $L$ light curves.
Those at $H$ and $K$ are intermediate between these. Fig.~2 shows the fit to
the $J$ data of a sinusoid with $P_0=575$ days and a long-term trend,
modelled as sinusoid with $P_1 = 12 \times P_0$; a small contribution is also
made by the first harmonics of the two periods, $P_0/2, P_1/2$. The data
stream is not long enough to determine the nature of the long-term trend and
the fact that it can be represented in this way is probably fortuitous.  

The 575 day period, which we take as the pulsation period, is well defined.
This is only a slight revision of the period of 580 days given by Le Bertre
(1992).  Evidently this revised period satisfies the early observations
(i.e. JD\,2447000-2448000) when the object was faint at $J$ and also the
later ones (JD\,2451600-2452600) when it was bright at $J$.  This apparent
constancy of the period suggests that there has been no significant changes
in the properties of the underlying star during the period of observation.
The amplitudes of the pulsation are slightly smaller at the brighter epoch
(by $\sim0.2$ mag at $J$, $\sim 0.4$ mag at $H$, $\sim 0.2$ mag at $K$ and
$\laeq 0.1$mag at $L$). In view of the known irregularities of Mira light
curves it is not clear whether these small changes have any important
significance.  However, they may be due to changes in the relative
contributions of the starlight and the dust emission to the total flux at
the early and late epochs (see below).

The near constancy of the mean $L$ brightness is striking compared with the
large changes at $J$ (a range of more than two magnitudes). There is,
however, a small increase in the brightness at $L$ (of about 0.1mag) between
the early (faint) and later (bright) epochs.

It is evident from Fig.~1 that the mean brightness of II\,Lup at $J$
underwent an extended minimum lasting from at least JD $\sim$ 2445800 to JD
$\sim$ 2451800. Despite the sparseness of the data at some epochs it is
clear that the $J$ variations during this period were complex and that the
curves shown in Fig. 2 provide only an approximate fit to the long term
variations. There was evidently a steep decline in mean $J$ brightness near
JD\,2446000. Between JD $\sim$ 2447000 and JD $\sim$ 2449200 there appears
to have been a further, slower, decline followed by a rise in brightness. A
relatively steep decline followed at JD $\sim$ 2449600. From JD $\sim$
2450500 to the end of the time period under discussion (JD\,2452500), the
object has been brightening at $J$.

It is of interest to compare the brightness variations of II\,Lup at $J$
with some theoretical models. There has been extensive work over a number of
years on the connection between pulsation, mass-loss and dust formation. 
See, for instance, the summaries and references to earlier work by two
particularly active groups in Sedlmayr \& Winters (2000), Fleischer et al.
(2000), Winters (2003), H\"{o}fner \& Dorfi (1997), Sandin \& H\"{o}fner
(2003), and Andersen et al. (2003). This work has centred on the formation
of shock waves and dust in the circumstellar environment. A paper by Winters
et al. (1994) makes predictions about the long term light curves expected
from models of this type. These predictions are sensitive to the C/O element
abundance ratio adopted in the models, as was particularly emphasized by
Fleischer et al. (2000). For relatively high values of the ratio ($\sim$
1.8) no long term modulation of the light curve is predicted. However, for
lower values (e.g. $\sim 1.3$) a long term periodicity is predicted, new
dust being formed not every pulsation cycle but on a longer, periodic, time
scale. Such a simple periodicity is not apparent in the behaviour of the $J$
brightness of II\,Lup just described. This result is similar to that
obtained by Whitelock et al. (1997) for a number of other carbon Miras.
These authors pointed out that the infrared light curves of R\,For were
complex and did not conform to the model predictions. R\,Vol and R\,Lep were
similar to R\,For.

There are some carbon variables, e.g. R\,Scl (Whitelock et al. 1997), which
may show long-term periodic modulation in their infrared light curves,
though this needs confirmation.  Note that the carbon variable V\,Hya with a
secondary (optical) period of $\sim 6160$ days (Mayall 1965, Knapp et al.
1999, Olivier et al. 2001) is thought to be in a binary orbit of this
period. However, it would appear that in general carbon Miras do not show
the long term periodic modulation predicted by some models. The long term
variations of these stars are much less regular. It may be noted that the
model of Winters et al. (1994) has been criticized by
Groenewegen (1997) in its application to CW\,Leo because it requires 
a different 
distance for the object to fit the 11 $\mu$m visibility curve 
from that needed to fit the spectral energy distribution.
We may conclude that
whilst such theoretical models may well form the basis for further advances,
they do not fit all the observations. The current models are based on
spherically symmetrical systems. In the next section we give reasons for
supposing that there were significant departures from spherical symmetry in
the case of the II\,Lup during the time period under discussion.

\subsection{Evidence for a Dust Cloud Model}
 In view of the apparent constancy of the pulsation period through the time
period of the observations discussed above, we assume in the following that
there has been no overall change in the underlying star.  The tentative
suggestion (Whitelock et al. 1997) that in the case of R\,For an apparent
change in pulsation period might be associated with a change in mass-loss
rate, needs further investigation.

Infrared speckle observations of II\,Lup were obtained by Lopez et al.
(1993) on JD\,2448434 when the object was in the obscuration minimum (see
Fig.~1 or 2).  These workers used their observations together with a
spherically symmetrical model to estimate the relative contributions of the
dust emission and starlight to the total flux at various wavelengths. In
view of the departure from spherical symmetry which we are suggesting in the
present paper, the numerical results derived by Lopez et al. have to be
treated with caution. Nevertheless it seems safe to assume from their work
that at this time almost all of the flux at $L$ came from the dust shell
whilst almost all of the flux at $J$ was starlight. The near constancy of
the mean $L$ flux throughout the period of our observations then suggests a
constant dust shell. On the other hand the variations at $J$ suggest changes
in the shell opacity in the line-of-sight. This evidently requires an
asymmetric model.

To roughly quantify these results we adopt the proportions of flux from the
star and the shell at different wavelengths as given by Lopez et al. Table 1
shows that (by chance) we have a $JHKL$ observation on the same day as their
observation. The object was then around phase 0.16 in the pulsation cycle.
Lopez et al. (see their fig.~6) interpret their results to show that at that
time there was a negligible contribution of heated dust in the shell to the
flux at $J$. Most of the $J$ flux was direct starlight though about 10
percent of the flux was scattered starlight. In contrast, at $L$ 80 percent
of the flux was dust emission and only 20 percent starlight.

Towards the end of
the time period under discussion, when the object was bright at $J$, it
was observed at almost exactly the same pulsation phase as that of the
speckle work (JD\,2452457 see Table~1). Call this epoch B and the epoch of the
speckle work A (these epochs are marked in Fig.~1). 
Suppose
the variation in $J$ between A and B is due to a change in dust absorption.
The speckle work indicates that at epoch A 15 percent
of the flux at $H$ was from heated dust. 
Adopting this value and assuming that
the flux from heated dust remained constant, it follows from the photometry
at A and B that the effective absorption law of the dust is 
$\Delta J_{*}/ \Delta H_{*}$ = 1.24, where the asterisk indicates the
stellar component of the flux\footnote{Without allowing for the dust
component at $H$ this ratio is 1.35}. This is similar to the effective
reddening law found for the carbon Mira R\,For (Feast et al. 1984) which
gives 1.33 for this ratio.  The uncertainty of this quantity is difficult to
estimate, but may be about 0.1 in the case of both variables. 
This agreement with R\,For, which has much bluer $J-L$ colours (see below),
suggests that the conclusions of Lopez et al. (1993), assigning the
bulk of the $J$ flux to the stellar contribution, are broadly correct.
In connection
with their discussion of circumstellar dust formation round carbon
variables, Andersen et al. (2003) consider three types of carbon-rich
particles using data taken from Rouleau \& Martin (1991) and J\"{a}ger et
al. (1998). The predicted values for the above ratio range from 1.36 to 1.85
for the three types of particle. The rough agreement with the values
actually observed is probably as good as can be expected in view of both
observational and theoretical uncertainties. The fact that the speckle work
indicates that dust emission contributes a large fraction (50 percent) of
the flux at $K$ at epoch A, makes it impractical to extend this type of
calculation to wavelengths longer than $H$.

If 80 percent of the flux at $L$ is from dust emission at epoch A then 
$L_{sh} = -0.15$, where the subscript ``sh" denotes shell, and 
$L_{*} = +1.36$.  
The work on R\,For (Feast et al. 1984) gave $\Delta J/ \Delta L \sim 4.0$ as
the effective reddening ratio of the circumstellar shell. This value is
uncertain as it might be affected by (an unknown) shell-emission component
at $L$.  But note that the range covered by the infrared colours of R\,For
used to determine the reddening law are bluer than those of II\,Lup (e.g. a
range in $(J-L)_{0}$ of 4.5 to 5.4 for R\,For compared with values for
II\,Lup at epochs A and B of 8.0 and 6.0). Thus the contribution of dust
emission to the $L$ flux is likely to be lower for R\,For than for II\,Lup.
The three dust models considered by Andersen et al. (2003) predict the ratio
$\Delta J/ \Delta L$ to be in the range from 3.2 to 5.8. If the shell flux
at $L$ has remained constant between epochs A and B we can use an adopted
ratio $\Delta J_{*}/ \Delta L_{*}$ to predict the brightening of the stellar
component of $L$, and hence the expected total brightness at $L$. For
adopted values of $\Delta J_{*}/ \Delta L_{*}$ of 3, 4, 5 and 6 one then
predicts the $L$ magnitudes at epoch B to be $-0.55, -0.51, -0.49$ and
$-0.47$. The observed value is $-0.49$. Considering the various
uncertainties there is no disagreement with this range of models. We may
conclude that the observations are consistent with the hypothesis of
constant emission from the shell and a varying stellar component between
epoch A and B due to variable dust absorption.

It may be noted that if we ascribed the increase in brightness at $J$ between
epochs A and B to an increase in the emission flux from the dust, then
the dust flux at $J$ would need to increase by about 6 magnitudes. A similar
increase at longer wavelengths is then anticipated unless the temperature
structure of the shell changed markedly. Any such change would have
to take into account that the pulsation amplitudes are little affected
by the rise in brightness. Although the $J$ flux at B would then be essentially
all from the dust flux whilst at A it is
essentially all starlight. In addition the total $L$ flux has to be
kept nearly
constant. We therefore discard this possibility.

We might anticipate that if the increase in brightness at $J$ between A and B
were due to the dispersal  of a spherical shell  of dust, then the
circumstellar dust emission would be reduced at the same time. An estimate
of this reduction could be made if we knew the intrinsic colours of the
underlying star. These are not known. However, the carbon Miras in the
LMC discussed by Feast et al. (1989) are all relatively bright optically
and 
$J-H$ is likely
to be only slightly affected by circumstellar reddening. 
These twenty Miras have
$\overline{J-H} = 1.23$ with a range of individual values from 0.70 to
1.67.  
The five carbon Miras with periods greater than 350 days have
$\overline{J-H} =1.39$ and a range 1.21 to 1.67. The longest period Mira in
this group has a period of 418 days and $J-H = 1.30$.  The above values need
to be corrected for the interstellar reddening of the LMC stars. This is
expected to be small ($E_{(J-H)} \sim 0.03$ from Feast et al. 1989). 
The optically bright (unobscured) carbon Miras in our Galaxy cover a
similar range in $(J-H)_{0}$ colours (Feast et al. 1982).
One
might expect the intrinsic colours of carbon Miras to increase with period.
This is certainly the case for oxygen Miras. For the purposes of
illustration we consider the consequences of adopting intrinsic colours
$(J-H)_{0}$ of 1.0 or 1.5 for II\,Lup.  We also adopt an interstellar
absorption of $A_{V} = 0.8$ mag from Groenewegen et al. (1998). It is then
straightforward to show that with 80 percent of the flux at $L$ from the
shell at epoch A, there would be a decrease in the total flux at $L$ between
A and B of 0.12 or 0.23 mag, assuming that the temperature of the dust
emission remained constant. These figures account for both the expected
decrease in the shell emission and the increase in starlight at $L$. In fact
the decrease in obscuration at $J$ between A and B is presumably due to to
the dust concerned moving away from the star. Thus the temperature of this
dust will decrease between A and B and the actual decrease in brightness at
$L$ will be greater than the figures just given.  However, we in fact see a
small increase ($\sim 0.1$ mag) in the total flux at $L$ between A and B
which (as discussed above) can plausibly be attributed to the increase in
direct starlight at the later epoch. These considerations suggest that we
are not dealing with the dispersal of a complete dust shell. But with the
clearing (due to expansion from the star) of a dust cloud of limited extent
and in the line-of-sight. Note that for simplicity we refer to a dust cloud.
However, in view of the relative complexity of the $J$ light curve, ejection
in the line-of-sight was presumably taking place over a considerable length
of time and at a variable rate.

\subsection {General Discussion}
 The conclusion reached above, that the long-term modulation of the light
curves of II\,Lup is due to the ejection of a dust cloud (or clouds) of
limited extent in the line-of-sight, leaving the circumstellar dust emission
virtually unchanged, is consistent with other data on carbon Miras. In the
case of R\,For, it was argued by Whitelock et al. (1997) that an obscuration
phase was due to a dust cloud which could not be a complete spherical shell.
These authors also point out that the bolometric absolute magnitude of
R\,Lep calculated from observations from the near- to the mid-infrared and
using the Hipparcos parallax of the object, is fainter than anticipated
theoretically.  This conclusion is not changed if one adopts the revised
parallax of R\,Lep recently published by Knapp et al. (2003). Whitelock et
al. (1997) note that this problem can be overcome if the circumstellar
emission from R\,Lep is from a non-uniform shell with higher than average
absorption in the line-of-sight.

One, extreme,
model for a non-spherical dust shell 
would be a circumstellar disc viewed edge-on. Disc models have been
suggested to explain polarization observations of the carbon variable RW\,LMi
(GL1403) (Cohen \& Schmidt 1982) which has some similarities to R\,For (see
Whitelock et al. 1997). Theoretical disc models have also been proposed 
(Dorfi \& H\"{o}fner 1996). Such a model might be appropriate for the 
unusual object V\,Hya (e.g. Dorfi \& H\"{o}fner 2000)
which is probably a binary (see above). However, it is not
certain that it is applicable to carbon Miras in general.  For instance if
the disc is optically thick to starlight we would expect significant 
changes in apparent bolometric luminosity between pole-on and edge-on views.
For a disc with total opening angle (measured at the centre of the star)
of 20 degrees this would amount to about two magnitudes. The scatter
of long period carbon Miras with dust shells in the LMC about a bolometric 
period-luminosity relation is less than one magnitude (Whitelock et al. 2003)
and much of this is likely to be due to the very poor 
temporal coverage of the stars
at mid-infrared wavelengths. 

In the sample of LMC carbon Miras studied by Whitelock et al. (2003) 
there are four
which show evidence of obscuration phases. These, together with their
$K-[12]$ colours 
(outside major obscuration)
are; IRAS\,05300--6651, 7.29; TRM\,72, 5.93; IRAS\,05009--6616, 6.28;
TRM\,88, 4.44. The colours order the objects 
according to the relative contributions
of the near- and mid-infrared to their bolometric fluxes. For the last three
objects the near-infrared data are sufficient to allow a comparison between
the bolometric magnitudes obtained using $JHKL$ values at the bright and faint
phases of the obscurations. 
For these three stars (in the order given above) the differences between
these values are only 0.1, 0.1 and 0.4 mag.
The $K-[12]$ colours of the last of these, TRM\,88,
are amongst the smallest of the long period sample of carbon Miras used
by Whitelock et al. to define a period-luminosity relation and the object has 
a large  obscuration range ($\Delta K \sim$ 1 mag).
Thus the conclusion drawn above regarding the relative narrowness of the
period-luminosity relation
is not affected by the way the bolometric magnitudes were calculated by
Whitelock et al.

The optically bright and relatively short period ($P < 420$ days) carbon Miras
in the LMC discussed by  Feast et al. (1989) fit tight period-luminosity
relations at $K$ and $M_{bol}$. Such stars thus appear to have thin shells
and their $K$ and $M_{bol}$ magnitudes are not seriously affected 
by circumstellar reddening. However, the Galactic carbon Mira, R\,Lep, 
mentioned above is an interesting intermediate case. With $K-[12] = 2.41$
its near infrared flux will contribute more to a derived $M_{bol}$ than in
the case of the obscured LMC Miras mentioned above. Thus higher than average
shell absorption in the line-of-sight could well lead to a considerable
underestimate of the bolometric luminosity as suggested by Whitelock et al.
(1997). II\,Lup, itself, with $K-[12] = 3.39$ may also be in this category. 

Further work on the obscured LMC stars might allow one to discriminate 
decisively
between a model with a disc of large vertical extent and a quasi-spherical
model. At present
the latter seems the more promising for most of the stars. Both models
require small scale variations due to dust clouds in order to explain
object like II\,Lup unless one invokes a precessing disc model. Even then
some non-uniform structure in the disc would be necessary to take into
account the complexity of the variations at $J$.

There are a number of other observations which point towards
non-uniform shell models.
High resolution observations show that the inner region of
the circumstellar material around CW\,Leo is highly structured  and variable
(e.g. Tuthill et al. 2000, Men'shchikov et al. 2001 and the summary in
Monnier et al. 2003),
whilst remarkable images of the outer region show complex arclet structures
(de Laverny  2003).
Furthermore CO observations show the shells of carbon variables  
(mostly classed as semi-regular) to be
clumpy (e.g. Olofsson et al. 1996, 1998). 
 
It is interesting to note that obscuration events appear to be relatively 
rare in single oxygen Miras (see Whitelock et al. 2000 and also 
Bedding et al. 2003). 
It is not clear whether or not
this difference between carbon and oxygen Miras could be due to the type of 
particle involved. Whitelock (1987, 2003)
has drawn attention to the fact that obscuration minima are relatively
frequent in symbiotic systems containing oxygen Miras. She suggests that
this might be due to the wind from the hot component in the system
producing inhomogeneities in the distribution of the dust.

The long term modulation of the light curves of carbon Miras by dust clouds
ejected in the line-of sight is similar in principle to the model for
the obscuration minima of R\,Coronae Borealis (RCB) variables, 
with the star itself left unchanged (a good example of this in the
case of the pulsating RCB star RY Sgr is given in Feast (1979) and
reproduced in Clayton (1996)).
If the $\rm C_{2}$ bands seen in emission during obscuration phases of
some carbon Miras (Lloyd Evans 1997) come from the outer atmosphere of these 
stars they could be explained as becoming visible when the main body of the
star was dimmed by a dust cloud. This is the model for the emission
spectrum seen in RCB stars during minima (Feast 1979). Whitelock et al. (1997)
note some
similarities between the long term variations of R\,For and RCB stars.  
Some differences
between the details of the Mira obscuration phases and those
of typical RCB stars, are, however, to
be expected in view of the major differences between the size and temperatures
of the stars involved, the order of magnitude difference in the expansion
velocities  of the 
circumstellar material, and possible differences in the effective reddening
law of the shell.

Understanding the formation of the dust is a formidable problem and it
remains to be seen whether the theoretical advances cited above can be
combined with an asymmetrical ejection (dust cloud) model. The basic cause
of asymmetrical mass ejection may be non-uniformities in the stellar
atmosphere. At least in the case of oxygen Miras, interferometry has
indicated departures from circular symmetry which might be due areas of
different brightness (and temperature) (e.g. Lattanzi et al. 1997), possibly
large convection shells.  It has also been suggested (Soker \& Clayton 1999)
that dust formation in both AGB and RCB stars takes place preferably above
cool magnetic spots. Alternatively, instabilities in the dust forming region
itself may lead to the formation of dust clouds (Woitke et al. 2000).

\section{Conclusions}
  Superposed on the pulsational brightness variations of II\,Lup at $J$
there is a long term, large amplitude variation. This is complex and, as in
the case of R\,For (Whitelock et al. 1997) and probably other stars, it is
not easily fitted by available models which predict long-term periodic
variations for mass-losing carbon Miras. These models are spherically
symmetrical. Combining our $JHKL$ data with the high resolution work of
Lopez et al. (1993) allows us to conclude that the effective reddening law
of the circumstellar material at $J$ and $H$ is similar to that found for
R\,For (Feast et al. 1984). The overall properties of the star itself and
the radiation from the circumstellar dust have remained practically constant
over the $\sim 18$ years of observation. The long-term variations at $J$ are
attributed to a dust cloud or clouds ejected in the line-of sight. Taken
together with other evidence the most likely model for objects of this kind
is a quasi-spherical dust shell, formed in material moving away from the
star, with small scale irregularities which can produce large variations in
the line-of-sight absorption as they form and disperse. The model is then
similar to that proposed for RCB variables which are also carbon-rich
(though hydrogen deficient). It is hoped that future interferometric work
will be able to test the model we propose for II\,Lup.

\section*{Acknowledgements}
This paper is based on observations made at the South African Astronomical
Observatory (SAAO). We are grateful to the following who each made some of
the observations: Brian Carter, Robin Catchpole, Dave Laney, Karen Pollard
and Greg Roberts.

\end{document}